\documentclass{easychair}

\usepackage{doc}

\usepackage{graphicx}

\usepackage{cite}
\usepackage{booktabs}
\usepackage{bm}
\usepackage{amsmath,amssymb,amsfonts}
\usepackage{tabularx}
\usepackage{algorithmic}
\usepackage{graphicx}
\usepackage{textcomp}
\usepackage{float}
\usepackage{overpic}
\usepackage{multirow} 
\usepackage{enumerate}
\usepackage{longtable}
\usepackage[caption=false]{subfig}
\usepackage[ruled,linesnumbered]{algorithm2e}
\usepackage{marvosym}

\newcommand\keywords[1]{\textbf{Keywords}: #1}

\usepackage{rotating}
\usepackage{pdflscape}

\title{SFPDML: Securer and Faster Privacy-Preserving Distributed Machine Learning Based on MKTFHE}

\author{
    Hongxiao Wang\inst{1} \and
    Zoe L. Jiang\inst{2,} \inst{3} \and
    Yanmin Zhao\inst{1} \and
    Siu-Ming Yiu\inst{1(\text{\Letter})} \and
    Peng Yang\inst{2} \and
    Man Chen\inst{4}  \and
    Zejiu Tan\inst{2} \and
    Bohan Jin\inst{2} 
}

\institute{University of Hong Kong, Hong Kong, China \\
\email{\{hxwang, ymzhao, smyiu\}@cs.hku.hk} \and
Harbin Institute of Technology, Shenzhen, Shenzhen, China\\
\email{zoeljiang@hit.edu.cn} \\
\email{\{stuyangpeng, 23s151118, 23s051024\}@stu.hit.edu.cn} \and
Peng Cheng Laboratory, Shenzhen, China\and
Shandong University, Jinan, China \\
\email{chenman19961121@gamil.com}
}
\authorrunning{}

\titlerunning{SFPDML: Securer and Faster Privacy-Preserving Distributed Machine Learning based on MKTFHE}

\begin{document}

\maketitle

\begin{abstract}
In recent years, distributed machine learning has garnered significant attention. However, privacy continues to be an unresolved issue within this field. Multi-key homomorphic encryption over torus (MKTFHE) is one of the promising candidates for addressing this concern. Nevertheless, there may be security risks in the decryption of MKTFHE. Moreover, to our best known, the latest works about MKTFHE only support Boolean operation and linear operation which cannot directly compute the non-linear function like Sigmoid. Therefore, it is still hard to perform common machine learning such as logistic regression and neural networks in high performance.

In this paper, we first discover a possible attack on the existing distributed decryption protocol for MKTFHE and subsequently introduce secret sharing to propose a securer one. Next, we design a new MKTFHE-friendly activation function via \emph{homogenizer} and \emph{compare quads}. Finally, we utilize them to implement logistic regression and neural network training in MKTFHE. Comparing the efficiency and accuracy between using Taylor polynomials of Sigmoid and our proposed function as an activation function, the experiments show that the efficiency of our function is 10 times higher than using 7-order Taylor polynomials straightly and the accuracy of the training model is similar to using a high-order polynomial as an activation function scheme.

\end{abstract}

	



%
%

\section{Introduction} \label{sect1}
In the big data era, it is necessary to transform centralized systems into distributed ones in machine learning tasks. However, these distributed systems lead to new challenges, and one of the most pressing is privacy~\cite{verbraeken2020survey, PMP4MLDS18}.

Privacy computing is a technique that enables data computation without any risk of information leakage. To outsource private computations, fully homomorphic encryption (FHE), a cryptographic tool, is employed. FHE is a unique form of encryption that allows users to perform computations on encrypted data without the need to first decrypt it. FHE can be divided into two categories: \emph{single-key} fully homomorphic encryption and \emph{multi-key} fully homomorphic encryption.

\emph{Single-key} FHE only allows a server to perform addition and multiplication on data encrypted by the same key. In contrast, \emph{multi-key} FHE (MKFHE)  proposed in~\cite{Lopez12} enables users to encrypt their own data under their own keys, but during the decryption of MKFHE, all secret keys of all participants are used. It prevents conspiracy between a user and a server to steal the data of other users.
 

In recent years, multi-key fully homomorphic encryption over the torus (MKTFHE) has attracted significant attention from researchers, particularly in the areas of evaluation and decryption algorithms. Chen et al.~\cite{MKTFHE} developed a library for implementing MKTFHE, focusing on an evaluation algorithm that takes a NAND gate as input. Subsequently, Jiang et al.~\cite{MKTFHE-op} expanded the evaluation algorithm to include arithmetic operators such as adders, subtracters, multipliers, and dividers, enabling linear multi-key homomorphic arithmetic evaluation in MKTFHE. However, the inability to evaluate non-linear operations, like the Sigmoid function, restricts the direct application of more complex machine learning schemes such as logistic regression and neural networks.


To overcome this limitation, we borrow the SecureML~\cite{mohassel2017secureml} idea of replacing the non-linear activation function with a piecewise function (MKTFHE-friendly activation function). During the evaluation of the activation function, operands are transformed into Booleans and evaluated using Boolean operations. However, implementing piecewise functions in the FHE scheme is challenging due to their discontinuous nature and the need for input ciphertext comparisons with each bound. Instead of the online interactive comparison utilized in SecureML, we propose a simpler method: We develop \emph{compare quads} to \emph{select} ciphertext $ct_c$ and $ct_d$ by comparing messages between ciphertext $ct_a$ and $ct_b$. This is achieved by subtracting $ct_a$ and $ct_b$, extracting the most significant bit (MSB) of the result, and using the MSB to \emph{select} the appropriate value through homomorphic evaluation of the equation $\mathsf{MSB}\land ct_c + \lnot \mathsf{MSB }\land ct_d$. Our \emph{compare quads} supports more complex branching programs and enables the implementation of the SecureML-like piecewise function with only two \emph{compare quads}.


Concerning the decryption algorithm, Chen et al.~\cite{MKTFHE} provided a naive decryption algorithm for the original MKTFHE, which requires all users' secret keys as input. In practical scenarios, access to others' secret keys during decryption should be restricted. To address this issue, Lee et al.~\cite{LeeP19} proposed a distributed decryption algorithm that separates the decryption process into two sub-algorithms: partial decryption and final decryption. Each user employs their secret key for partial decryption, ensuring that no user has access to others' secret keys. However, we discover some \emph{possible attacks} on the distributed decryption algorithm, described in Appendix~\ref{Apx_1}.


\emph{Why the previous decryption is not secure}. In brief, the existing MKTFHE scheme leaks information about user $u_{i}$'s secret key $s_{i}$ when provided with ciphertext and partial decryption. Suppose an MKTLWE ciphertext $({\bf a}_{1},\ldots, {\bf a}_{k},b)$, with $b=\frac{1}{4}m-\sum^{k}_{i=1} \langle {\bf a}_{i}, {\bf s}_{i} \rangle + e$, $k$ as the number of users, $m$ as a one-bit message, and $e$ as an error to randomize $b$. With the partial decryption $p_{i}=b+\langle {\bf a}_{i},{\bf s}_{i} \rangle$ and ciphertext $b$, one can obtain $\langle {\bf a}_{i},{\bf s}_{i} \rangle$ via their subtraction which may leak information about ${\bf s}_{i}$. Besides, an external adversary obtaining all partial decryption results can finish the final multi-key encryption results alone by computing $\sum_{i=1}^{k}p_i-\left(k-1\right)b$. 

To address these security concerns, we introduce additive secret sharing to protect partial decryption and facilitate final decryption. Each participant performs partial decryption using their secret key and shares the result with decryption parties. Secret sharing is then employed to complete the final decryption, preventing internal adversaries from accessing others' secret keys and external adversaries from obtaining partial decryption results. Since only additional operations are involved, decryption parties do not need to interact, making our decryption protocol secure against both internal and external adversaries while maintaining a similar cost to the original algorithm.

Finally, we combine these advancements to propose our Secure and Faster Privacy-Preserving Distributed Machine Learning (SFPDML) scheme, applying it to train and predict logistic regression models and neural networks using the Iris dataset.

Our contributions can be summarized as follows:
\begin{enumerate}[1.]
	\item 
	We develop a secure distributed decryption protocol for MKTFHE by introducing a secret sharing scheme, addressing the information leakage problem. We define our security goal for MKTFHE against possible static adversaries, then prove the correctness and analyze the security of our protocol. Experimental results demonstrate acceptable performance for the overall scheme and achieve similar performance to the original algorithm when the number of participants increases.
	\item 
    We apply our SFPDML scheme to train and evaluate logistic regression and neural network models. We design a \emph{homogenizer} to modify the bit length of operands, allowing the use of fewer-bit operators to reduce operation time. To implement a new MKTFHE-friendly activation function, we develop \emph{compare quads} for comparing and selecting input ciphertext. Experimental results show that our function's efficiency is 10 times higher than using 7-order Taylor polynomials directly, and the accuracy of the trained model is similar to that of high-order polynomial activation function schemes.
\end{enumerate}

\section{Related Work}
There are numerous works on privacy-preserving machine learning prediction~\cite{CryptoNets16, BourseMMP18, BoemerLCW19, TianNYY21, CHET19} and training~\cite{ChenGHHJLL18, KimS0XJ18, CheonKKS18}. These solutions are based on single-key FHE which cannot support the data participants using different secret keys to encrypt their own data. The prediction can support more complex models such as logistic regression and even neural networks in the second level, but the training only focuses on simpler models such as logistic regression in the hour level or higher. Besides, we also note that there are other approaches based on MPC, e.g.~\cite{mohassel2017secureml, makri2017pics} and compared with the above works based on FHE, the performances of the solutions based on MPC are very impressive. But they need interactivity between the data participants and the computation parties which may lead to many problems such as network latency or high bandwidth usage. Considering the above downsides, we focus on FHE, especially multi-key FHE.

The concept of multi-key fully homomorphic encryption was first proposed by Lopez et al.~\cite{Lopez12}, which is intended to apply to on-the-fly multiparty computation based on NTRU. Then first LWE-based MKFHE was constructed by Clear et al.~\cite{Clear15}, and later, was improved by Mukherjee and Wichs~\cite{Mukherjee16}. These schemes are single-hop MKFHE schemes, which means all the participants must be known in advance. The multi-hop MKFHE schemes were proposed by Peikert et al.~\cite{Peikert16} and Brakerski et al.~\cite{Bra16}, but their schemes are impractical and without implementation. 

The first implementation of the MKFHE scheme was achieved by Chen et al. ~\cite{MKTFHE}, named MKTFHE which is the variant of TFHE ~\cite{ChillottiGGI16, ChillottiGGI17, ChillottiGGI20}. Their scheme only provided a bootstrapped NAND gate to evaluate. Then, Lee and Park~\cite{LeeP19} first formalized the distributed decryption for MKFHE and improved the decryption part of MKTFHE, but a passive adversary still can recover the decryption result through the partial results. Then Jiang et al.~\cite{MKTFHE-op} designed other bootstrapped gates, utilized them to build arithmetic operators including adder, subtractor, multiplier, and divider, and then implemented a privacy-preserving linear regression in the GD method.

However, only using arithmetic operators cannot directly compute the non-linear activation function like the Sigmoid function. So, there is still a gap between MKTFHE and the implementation of more complex privacy-preserving machine learning such as logistic regression and neural networks.

\section{Preliminaries}
\textbf{Notation}: 
In the rest of this paper, $\mathbb{R} $ denotes the real numbers, $\mathbb{Z}$ denotes the integers, and $\mathbb{T}$ indicates $ \mathbb{R}/\mathbb{Z}$, the torus of real numbers modulo $1$. We use TLWE to denote the (scalar) binary Learning With Error problem over Tours, and TRLWE for the ring mode. 
We define $params$ as the parameter set in TFHE, $ mkparams $ in MKTFHE, and our scheme. Besides, $ k$ is used to represent the number of participants in MKTFHE, $l$ is used to represent the bit length of a message or ciphertext, and $ct[l]$ is denoted as $l$-bit ciphertext in MSB order. Then we use bold letters, e.g. $\bm{a}$, to denote vector and use $\langle \bm{a},\bm{b}\rangle $ to represent the inner product between vector $\bm{a} $ and vector $\bm{b}$.

\subsection{Multi-key Fully Homomorphic Encryption over Torus}\label{AA}
MKTFHE scheme is the multi-key version of TFHE scheme~\cite{ChillottiGGI16, ChillottiGGI17, ChillottiGGI20}. 
In the MKTFHE scheme, the ciphertext length increases linearly with the number of users, and a homomorphic NAND gate with bootstrapping is given. The MKTFHE scheme is comprised of the following algorithms:
\begin{itemize}
\item $mkparams\gets \mathsf{MKTFHE.SETUP}\left(1^\lambda\right)$: Take a security parameter $\lambda$ as input and output the public parameter set $mkparams$.
\item $\{sk_i,pk_i\}\gets \mathsf{MKTFHE.KEYGEN} \left(mkparams\right)$: Take the $mkparams$ as input, and output secret key $sk_i$ and public key $pk_i$ for a single participant $i$.
\item $ct\gets \mathsf{MKTFHE.ENC}\left(\mu\right)$: Encrypt an input bit $\mu\in\{0,1\}$ and output a TLWE ciphertext with the scaling factor $\frac{1}{4}$. The output ciphertext $ct=\left(\bm{a},b\right)\in\mathbb{T}^{n+1}$, satisfing $ b+\langle \bm{a},\bm{s}\rangle \approx\frac{1}{4}\mu$.
\item $\mu\gets \mathsf{MKTFHE.DEC}\left(ct,\{sk_i\}_{i\in[k]}\right)$: Input a TLWE ciphertext $ct=(\bm{a_i},\cdots,$ $\bm{a_k},b)\in\mathbb{T}^{kn+1}$ and a set of secret keys $\{sk_i\}_{i\in[k]}$, and output the message $\mu\in\{0,1\}$ which satisfies $b+\sum_{i=1}^{k}{\langle \bm {a_i},sk_i\rangle }\approx\frac{1}{4}\mu\left(\mod1\right)$.
\item $ct\gets \mathsf{MKTFHE.NAND}\left(ct_1,ct_2\right)$: Input two TLWE ciphertext $ct_1=\mathsf{MKTFHE.}$ $\mathsf{ENC}\left(\mu_1\right)\in\mathbb{T}^{n+1}$, $ct_2=\mathsf{MKTFHE.ENC}\left(\mu_2\right)\in\mathbb{T}^{n+1}$, where $ct_1$, $ct_2$ can be constructed by different participants respectively, and output the multi-key ciphertext result $ct=\mathsf{MKTFHE.ENC}\left(\mu_1\oplus\mu_2\right)\in\mathbb{T}^{kn+1}$:
	\begin{itemize}
	\item[-] Extend $ct_1$ and $ct_2$ to $ct_1^\prime$ and $ct_2^\prime$ to make them encrypted under the multi-key $\{sk_i\}_{i\in[k]}$ by putting zero in the empty extending slots.
	\item[-] Evaluate $\mathsf{GB}\left(\left(0,\cdots,0,\frac{5}{8}\right)-ct_1^\prime-ct_2^\prime\right)$ and return the result.
	\end{itemize}
\end{itemize}

For the part $\mathsf{GB}$, we will not discuss it in this paper and refer to the original paper. And we call the evaluated ciphertext as multi-key ciphertext whose dimension is the number of participants in the MKTFHE scheme.

\subsection{Distributed Decryption}
The decryption algorithm of the existing MKTFHE is a single decryptor case, that is, the decryptor holds a set of secret keys $\{sk_i\}_{i\in[k]}$ of all participants. However, in practical use, for security reasons, the decryptor should not hold any secret key $sk_i$ of participants. Therefore, distributed decryption which involves all participants jointly decrypting a multi-key ciphertext is more practical. The most common distributed decryption for MKFHE~\cite{mukherjee2016two} has been defined below: 
\begin{itemize}
	\item $p_i\gets \mathsf{PartDec}\left(ct,sk_i\right)$: Input a multi-key ciphertext $ct$ under a set of secret keys $\{sk_i\}_{i\in[k]}$ of all participants, and the $i$-th secret key $sk_i$, output a partial decryption result $p_i$;
	\item $m\gets \mathsf{FinDec}\left(p_1,\cdots,p_k\right)$: Input a set of partial decryption results $\{p_i\}_{i\in[k]}$ of all participants and output the plaintext of the multi-key ciphertext.
\end{itemize}

\subsection{Homomorphic Gates and Operators Based on MKTFHE}
Based on the TFHE scheme and its multi-key variant, Jiang et al.~\cite{MKTFHE-op}. designed other binary gates with the same efficiency as NAND gates in MKTFHE including AND, OR, NOT, etc., and used their designed binary gates to implement the fixed $k$-bit complement arithmetic operators, so that addition, subtraction, multiplication, and division of both positive and negative numbers can be evaluated in MKTFHE. 
The definition of operators in MKTFHE is as follows:

\begin{itemize}
	\item $ct\left[l\right]\gets \mathsf{MKADD}\left(ct_1\left[l\right],ct_2\left[l\right]\right)$: Input two $l$-bit  TLWE ciphertexts $ct_1\left[l\right]$ and $ct_2\left[l\right]$ and output a $l$-bit MKTLWE ciphertext $ct\left[l\right]$.
	\item $ct\left[l\right]\gets \mathsf{MKSUB}\left(ct_1\left[l\right],ct_2\left[l\right]\right)$: Input two $l$-bit  TLWE ciphertexts $ct_1\left[l\right]$ and $ct_2\left[l\right]$ and output a $l$-bit MKTLWE ciphertext $ct\left[l\right]$.
	\item $ct\left[2l\right]\gets \mathsf{MKMUL}\left(ct_1\left[l\right],ct_2\left[l\right]\right)$: Input two $l$-bit  TLWE ciphertexts $ct_1\left[l\right]$ and $ct_2\left[l\right]$ and output a $2l$-bit  MKTLWE ciphertext $ct\left[2l\right]$.
	\item $ct\left[l\right]\gets \mathsf{MKDIV}\left(ct_1\left[2l\right],ct_2\left[l\right]\right)$: Input a $2l$-bit  TLWE ciphertext $ct_1\left[2l\right]$ and an $l$-bit  TLWE ciphertext $ct_2\left[l\right]$ and output an $l$-bit length TLWE ciphertext $ct\left[l\right]$.
\end{itemize}

Note that the input of the gate circuit is a single-bit ciphertext, while the input of the operator is a multi-bit ciphertext. Therefore, before inputting a multi-bit integer, firstly encode it as complement, and then encrypt it by bit. In addition, the bit-number of the multiplication and division input data and output data are different, which is prone to data overflow or the bits of the input data do not match the operator.	

\subsection{Secret Sharing Based on Arithmetic Circuit}
The secret sharing protocol on the arithmetic circuit is carried out on a finite field. In the secure 2-party computation, the $l$-bit value $x$ is shared by the participants into two elements on $\mathbb{Z}_{2^l}$ ring and sned the two elements to two computing parties $P_0$ and $P_1$ respectively. Make $\left[x\right]_i^A$ represents the sub secret owned by the computing party $P_i$, and the superscript $A$ represents the secret share on the arithmetic circuit. Secret sharing on arithmetic circuits is in $\mathbb{Z}_{2^l}$ which satisfies $\left[x\right]_0^A+\left[x\right]_1^A=x \mod{ 2}^l$ where $\left[x\right]_0^A$, $\left[x\right]_1^A\in\mathbb{Z}_{2^l}$. So, the participants can share and reconstruct the secret $x$ by the following algorithms:
\begin{itemize}
	\item $\{\left[x\right]_0^A,\left[x\right]_1^A\}\gets Share^A\left(x\right)$: Input an $l$-bit secret $x$, randomly choose $r\in\mathbb{Z}_{2^l}$, set $\left[x\right]_0^A=x-r$, $\left[x\right]_1^A=r$ and then output $\left[x\right]_0^A,\left[x\right]_1^A$.
	\item $x\gets Rec^A\left(\left[x\right]_0^A,\left[x\right]_1^A\right)$: Input $\left[x\right]_0^A$,$\left[x\right]_1^A$, compute $\left[x\right]_0^A+\left[x\right]_1^A$, then output the result.
\end{itemize}

In this protocol, if one party does not abide by the rules and sends the wrong value during the final reconstruction, the honest party cannot reconstruct the secret, while the fraudulent party can reconstruct the real secret. Therefore, this agreement is semi-honest and the participants need to abide by the rules of the agreement. 

Besides, the addition operation of this protocol is free and both computing parties can directly perform the calculation locally which follows below:
\begin{itemize}
	\item ${{\{\left[z\right]}_0^A,\left[z\right]_1^A}\}\gets {Ad d}^A\left(\left[x\right]_0^A,\left[x\right]_1^A,\left[y\right]_0^A,\left[y\right]_1^A\right)$: Input the secret shares of x, y, compute ${\left[z\right]_0^A=\left[x\right]}_0^A+\left[y\right]_0^A$, $\left[z\right]_1^A=\left[x\right]_1^A+\left[y\right]_1^A$, and output $\left[z\right]_0^A$,$\left[z\right]_1^A$.
\end{itemize}

\subsection{Machine Learning}\label{MLP}

\subsubsection{Logistic Regression}
Logistic regression is a generalized linear regression model. It is a classical method to solve the binary classification problem by using the activation function. In the traditional logistic regression, the activation function is defined as a Sigmoid function, function $f\left(x\right)=\frac{1}{1+e^{-x}}$. 
The Batch Gradient Descent (BGD) method for logistic regression updates the coefficients in each iteration as follows:
\begin{equation}
	z=\theta_0+\theta_1x_1+\cdots+\theta_nx_n=\theta^Tx
	\nonumber
\end{equation}
\begin{equation}
	h_\theta\left(x\right)=f\left(\theta^Tx\right)=\frac{1}{1+e^{-\theta^Tx}}
	\nonumber
\end{equation}
\begin{equation}
	\theta_j=\theta_j-\alpha\frac{1}{m}\sum_{i=1} n\left(h_\theta\left(x_i\right)-y_i\right)x_i^j.
	\nonumber
\end{equation}
The phase to calculate the predicted output $h_\theta\left(x_i\right)$ is called \emph{forward propagation}, and the phase to calculate the gradient $\alpha\frac{1}{m}\sum_{i=1} n\left(h_\theta\left(x_i\right)-y_i\right)x_i^j$ is called \emph{backward propagation}.




\subsubsection{Neural Networks}
Neural networks are a more generalized regression model compared to logistic regression to learn more complex relationships between high dimensional input data and multiple output labels. 
Traditional activation functions are like the Sigmoid function $f\left(x\right)$ or RELU function.

Standard error Back Propagation (BP) neural networks can be trained by  Gradient
Descent with Momentum (GDM) method so that the coefficient convergence will be faster than BGD and more stable than Stochastic Gradient Descent (SGD). By the chain rule, the coefficients are updated as follows:
$$ o_i=f(\sum_{i=1}^m w_{ji}x_i^k),{\hat{y}}_j^k=\sum_{i=1}^{n}{o_iv_{ji}},$$
$$ \Delta v_{ji}^{(n+1)}=\beta_1 \Delta_1v_{ji}^{(n)}+(1-\beta_1)(\hat{y}_j^k-y_j^k)o_i, $$
$$ \Delta w_{ij}^{\left(n+1\right)}=\beta_2\Delta w_{ij}^{\left(n\right)}+\left(1-\beta_2\right)o_i\left(1-o_i\right)x_j^k\sum_{j=1}^{p}{\left({\hat{y}}_j^k-y_j^k\right)v_{ji}}, $$
$$ v_{ji}^{\left(n+1\right)}=v_{ji}^{\left(n\right)}-\alpha_1\Delta v_{ji}^{\left(n+1\right)},{w_{ij}^{\left(n+1\right)}=w_{ij}^{\left(n\right)}-\alpha_2\Delta w}_{ij}^{\left(n+1\right)}, $$

\section{Distributed Decryption Protocol}
\subsection{Our Security Goal}
In the MKFHE scheme, participants generally generate their own secret keys independently, encrypt their own data by their own secret key, and decrypt the multi-key ciphertext by all secret keys jointly. Therefore, our security goal of MKFHE decryption is to protect individual single-key encrypted messages and common multi-key encrypted messages. 


As mentioned in Section~\ref{sect1}, we can observe that there are at least two kinds of static (passive) adversaries: an internal adversary and an external adversary. The internal adversary is one of the participants in the scheme, but the external adversary is not. Both adversaries want to know any information about each participant’s message and the external adversary also hopes to obtain the computation result. 

We define the security using the framework of Universal Composition (UC) ~\cite{canetti2001universally}. To simplify, we take both of them together as \emph{semi-honest} adversaries $\mathcal{A}$. And we assume the adversary $\mathcal{A}$ can corrupt any subset of the participants and servers (at least two participants and the server is uncorrupted) which only the data of the corrupted participants but nothing else about the remaining honest participants' data can be learned by the adversary $\mathcal{A}$. 

\subsection{Our Distributed Decryption Protocol for MKTFHE}
We already know that MKFHE is IND-CPA secure~\cite{cryptoeprint:2018:1156}, so the multi-key ciphertext is guaranteed to be secure for both adversaries and we force the protection of single-key ciphertext. Thanks to the technique of two-party secret sharing, no one can learn the whole single-key ciphertext, even the participants can only obtain their own part.

\begin{figure}[htbp]
	\centering
	\includegraphics[width=0.6\linewidth]{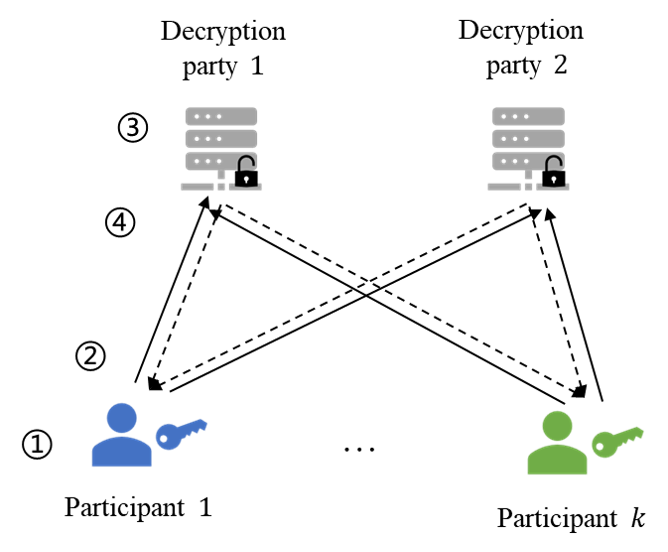}
	\caption{Our proposed distributed decryption protocol}	
	\label{DD_fig}
\end{figure}

Denote the multi-key ciphertext by $\hat{ct}=\left(\bm{a_1},\ldots,\bm{a_k},b\right)\in\mathbb{T}^{kn+1}$, which satisfies $b=\frac{1}{4}m-\sum_{j=1}^{k}{\langle \bm{a_j},\bm{s_j}\rangle }+e(\mod 1)$~\cite{MKTFHE}, where $m\in\{0, 1\}$ is the plaintext after decryption of the ciphertext, $k$ is the number of participants, $s_j$ is the private key of the $j$-th participant. The secure distributed decryption algorithm based on MKTFHE is defined as follows:
\begin{itemize}
	\item[-] Partial decryption algorithm: 

	$p_i\gets \mathsf{Part\_Dec}(\hat{ct}, s_i)$: The input is the multi-key ciphertext $\hat{ct}$ and the secret key $s_i$ of the $i$-th participant. The output $p_i$ is the partial decryption result of the $i$-th participant, and the computation is $p_i=b+\langle a_i,s_i\rangle $
	\item[-] Final decryption algorithm: 
	
	$\frac{1}{4}m+\bar{e}\gets \mathsf{Fin\_Dec}({\{p_i\}}_{i\in{k}})$: The input ${\{p_i\}}_{i\in{k}}$ is the partial decryption result of all participants, and the output is plaintext with noise before rounding, and the computation is $\frac{1}{4}m+\bar{e}=\sum_{i=1}^{k}p_i-(k-1)b$	
\end{itemize}
The protocol can be divided into four steps, as shown in Fig~\ref{DD_fig}
\begin{enumerate}[Step 1]
	\item Partial decryption: each participant $P_1,...,P_k$ uses their own secret key $s_i$ to compute partial decryption $\mathsf{Part\_Dec}(\hat{ct}, s_i)$ to obtain their personal partial decryption results $p_i$;
	\item Secret sharing: each participant runs the secret share algorithm for its own partial decryption result $p_i$. and get secret shares$\left[p_i\right]_0^A$ and $\left[p_i\right]_1^A$, and send them to the cloud server and the decryption party respectively;
	\item Offline computing: the cloud server and the decryption party receive secret shares received from participants, and then offline compute final decryption $\mathsf{Fin\_Dec}({\{\left[p_i\right]_0^A\}}_{i\in{k}})$ and $\mathsf{Fin\_Dec}({\{\left[p_i\right]_1^A\}}_{i\in{k}})$ respectively, so as to obtain the secret share of final decryption result $\left[\frac{1}{4}m+\bar{e}\right]_0^A$ and $\left[\frac{1}{4}m+\bar{e}\right]_1^A$;
	\item Secret reconstruction: the decryption party $DP_1$, $DP_2$ send the secret share of the final decryption result $\left[\frac{1}{4}m+\bar{e}\right]_0^A$ and $\left[\frac{1}{4}m+\bar{e}\right]_1^A$ to participants and the participants use the secret recovery protocol to reconstruct the final decryption result and obtain the final decryption result $\frac{1}{4}m+\bar{e}$.
\end{enumerate}

\subsubsection{Correctness Proof}
The correctness of the protocol follows:
\begin{equation}
	\begin{split}
&\left(\sum_{i=1}^{k}\left[p_i\right]_0^A-\left(k-1\right)\left[b\right]_0^A\right)+\left(\sum_{i=1}^{k}\left[p_i\right]_1^A-\left(k-1\right)\left[b\right]_1^A\right) \\
		&=\left(\sum_{i=1}^{k}\left[p_i\right]_0^A+\sum_{i=1}^{k}\left[p_i\right]_1^A\right)-\left(\left(k-1\right)\left[b\right]_0^A+\left(k-1\right)\left[b\right]_1^A\right)\\
		&=\sum_{i=1}^{k}p_i-\left(k-1\right)b=\sum_{i=1}^{k}\left(b+\langle a,s_i\rangle \right)-\left(k-1\right)b\\
		&=b+\sum_{i=1}^{k}{\langle a,s_i\rangle }=\frac{1}{4}m+e
		\nonumber
	\end{split}
\end{equation}
If the error $e$ is less than $\frac{1}{8}$, the decryption will work correctly.

\subsubsection{Security Proof}

In the UC framework, security is defined by comparing the \emph{real} world and \emph{ideal} world. The \emph{real} world is involved in the protocol, adversary $\mathcal{A}$, and honest participants. And the \emph{ideal} world includes the trusted party to represent the protocol, the simulator $\mathcal{S}$ to simulate the \emph{ideal} world, and honest participants. If the view of \emph{real} world and \emph{ideal} world is undistinguished, the protocol is secure. 

We consider security in the semi-honest model in which all participants and servers follow the protocol exactly. We assume that the two servers are non-colluding. We choose the adversary $\mathcal{A}$ who corrupts a server $DP_1$ and all but two of the participants $\{P_1, ..., P_{k-2}\}$ as an example that can cover all scenarios in our security goal. Simulator $\mathcal{S}$ is to simulate the above in \emph{ideal} world which submits the partial decryption of the participants and receives the final decryption from the trusted party. 

During the simulating, on behalf of the honest participants $\mathcal{S}$ sends a randomized partial decryption share $[p_i]_0^A$, $[p_i]_1^A$  in $\mathbb{Z}_{2^l}$ to $DP_1$ and $DP_2$. This is the only phase where participants are involved. Then each server evaluates independently until the final decryption is recovered. 

We can briefly argue that the view of the \emph{real} world and \emph{ideal} world is indistinguishable because of the security of the arithmetic secret sharing. The share of partial decryption is generated by participants randomly. Particularly, all messages sent, received, and reconstructed in our protocol are generated using uniformly random shares in both \emph{real} world our protocol involved and \emph{ideal} world simulator simulated, so the view of both identically distributed concludes our argument. 

\section{Privacy-Preserving Distributed Machine Learning}
\subsection{Pre-Work}
\subsubsection{Extract Sign Bit}
Thanks to the encryption and evaluation of MKTFHE bit by bit, we can easily extract any bit in a multi-bit ciphertext. In complement coding, the highest bit can represent the sign of the operand, which is that the highest bit is 0 and the operand is positive, and the highest bit is 1 and the operand is negative. Therefore, we can use this property to extract the sign bit of any ciphertext. The operation of extracting the sign bit is defined as follows:
\begin{itemize}
	\item $ct_{sign}\gets \mathsf{Extract\_Sign}\left(ct\left[l\right]\right)$: Input a $l$-bit length TLWE ciphertext $ct\left[l\right]$, and output the sign bit $ct_{sign}$ of $ct\left[l\right].$
\end{itemize}

\subsubsection{Cut off and Expand}
Considering the bit-length of the existing arithmetic operators in MKTFHE is predefined, and the larger the bit-length, the more time consumed. Therefore, flexibly adjusting the bit-number of ciphertext and selecting the less bit-length arithmetic operators can improve the efficiency and accuracy of the overall scheme of machine learning based on MKTFHE.

We continue to use the encoding method in MKTFHE arithmetic operators~\cite{MKTFHE-op} which is to use complement to encode both positive and negative integers. When the large-bit of ciphertext operands is put into the little-bit arithmetic operators, the operands will automatically cut off the rest bits of the input ciphertext operands, in other words, only the part of data with the same bit-number as the arithmetic operators will be calculated. For example, we can get a 16-bit ciphertext product from an 8-bit multiplier with a couple of 8-bit ciphertext operands input. And if this 16-bit ciphertext product doesn’t overflow 8-bit size, it can be put into the next 8-bit arithmetic operator directly with a little time of computation. But if the 16-bit ciphertext product overflows the size of 8-bit, it must be put into the 16-bit arithmetic operator in the next calculation with more time consumed, and its corresponding 8-bit operand must be expanded from 8-bit to 16-bit. Until the operand recovers the 8-bit size by subsequent operations such as division, we can continue to use the less bit-number operator.

\begin{figure}
  \begin{minipage}[t]{0.5\linewidth}
    \centering
    \includegraphics[width=0.8\linewidth]{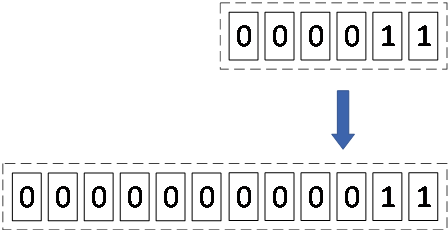}
    \caption{Expand the ciphertext}
    \label{HO_fig}
  \end{minipage}%
  \begin{minipage}[t]{0.5\linewidth}
    \centering
    \includegraphics[width=1\linewidth]{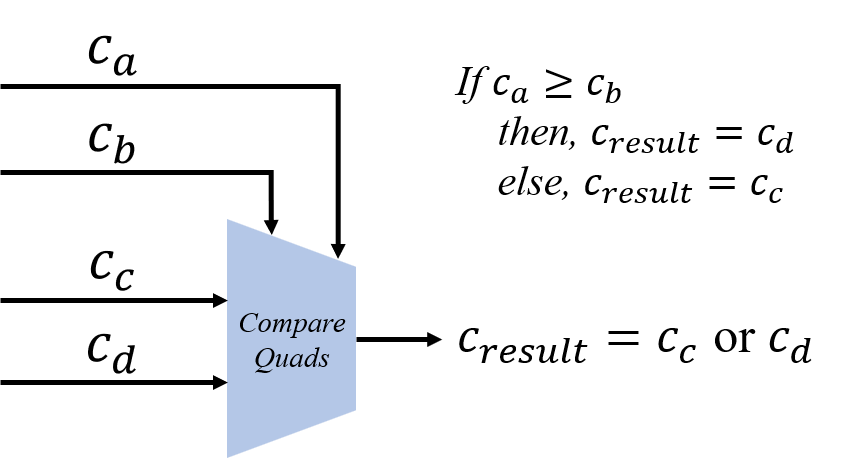}
    \caption{Select a ciphertext}
    \label{ACFC_fig}  
\end{minipage}
\end{figure}

In order to flexibly expand the bit-number of the operand and keep its sign, we design and implement a device named \emph{homogenizer}, as shown in Fig.~\ref{HO_fig}, and the specific design is as follows:

MKTFHE does not allow the ciphertext to be copied directly (it is considered unsafe), so we use the trivial $\mathsf{TLWE}(0)$ and the sign bit $ct_{sign}$ of the original small-bit ciphertext operand to calculate $\mathsf{MKAND}\left(\mathsf{TLWE}\left(0\right),ct_{sign}\right)$, and fill the ciphertext results into the high bits. The operation of expanding is defined as follows:
\begin{itemize}
	\item $ct\left[l^\prime\right]\gets \mathsf{Homogenizer}\left(ct\left[l\right],l^\prime\right)$: Input the $l$-bit length TLWE ciphertext $ct\left[l\right]$ and the bit-length $l^\prime$, output the $l^\prime$-bit length ciphertext $ct\left[l^\prime\right]$, the plaintext of which is the same as $ct\left[l\right].$
\end{itemize}

\subsubsection{Compare}\label{com_qua}
In the practical machine learning scheme, the comparison operation is usually required, but the existing MKTFHE scheme cannot support the comparison operation without decryption. We believe that the comparison can be divided into two categories. One needs to know the comparison results, such as the millionaire problem, and the other is to determine the next calculation through the comparison result, which is similar to branch selection. At present, the comparison in the machine learning scheme is mainly the second category. Therefore, we utilize the Boolean operation and arithmetic operations in MKTFHE to design and implement the basic elements of the comparison operation, named \emph{compare quads}, which is used to pick one from two ciphertext operands based on the results of comparison between the other two ciphertext operands, as shown in Fig.~\ref{ACFC_fig}. And the details of the \emph{compare quads} are described in Algorithm~\ref{alg1}. The operation of comparison is defined as follows:
\begin{itemize}
	\item $c_{result}\left[l\right]\gets \mathsf{Compare\_Quads}\left(c_a\left[l\right],c_b\left[l\right],c_c\left[l\right],c_d\left[l\right]\right)$: Input four $l$-bit length ciphertext $c_a\left[l\right]$, $c_b\left[l\right]$, $c_c\left[l\right]$, $c_d\left[l\right]$, if the plaintext of $c_a\left[l\right]$ is bigger than $c_b\left[l\right]$, then output $c_d\left[l\right]$, else output $c_c\left[l\right]$.
\end{itemize}



\begin{algorithm*}[t]
	\caption{\emph{Compare quads}}
	\label{alg1}
	\KwIn{MKTFHE parameter set $mkparams$, four $l$-bit ciphertext ${ c}_a\left[l\right]=\mathsf{MKENC}\left(a,l\right)$, $c_b\left[l\right]=\mathsf{MKENC}\left(b,l\right)$, $c_c\left[l\right]=\mathsf{MKENC}\left(c,l\right)$, $c_d\left[l\right]=\mathsf{MKENC}\left(d,l\right)$, and the public keys of all participants $\{pk_i\}_k$}
	\KwOut{$l$-bit ciphertext $c_{result}\left[l\right]=\mathsf{MKENC}\left(\left(a\geq\ b\right):d:c,l\right)$}  
	\BlankLine
    Compute $ \mathsf{MKSUB}\left(c_a\left[l\right],c_b\left[l\right]\right)$ to obtain $c_a\left[l\right]-c_b\left[l\right]$\\
    Compute $\mathsf{Extract\_Sign}\left(c_a\left[l\right]-c_b\left[l\right]\right)$
    to extract the sign bit $c_{sign}$\\
    Compute $\mathsf{MKNOT}\left(c_{sign}\right)$ to reverse the sign bit in order to obtain $\lnot c_{sign}$ \\
    Compute $\mathsf{MKAND}\left(c_{sign}, c_c\left[t\right]\right)$,  $\mathsf{MKAND}\left(\lnot c_{sign}, c_d\left[t\right]\right)$ to obtain $c_{sign}\land c_c\left[n\right]$, $\lnot c_{sign}\land c_d\left[n\right]$ which $t$ is from 1 to $n$ and can compute in parallel\\
    Compute $\mathsf{MKADD}(c_{sign}\land c_c\left[l\right],\lnot c_{sign}\land c_d\left[l\right])$ to obtain $c_{sign}\land c_c\left[n\right]+\lnot c_{sign}\land c_d\left[n\right]$ \\
    Return $c_{sign}\land c_c\left[n\right]+\lnot c_{sign}\land c_d\left[n\right]$
\end{algorithm*}

\subsection{MKTFHE Friendly Activation Function}
At present, the existing MKTFHE only supports integer linear operation and Boolean operation, so how to compute the Sigmoid function in logical regression and neural networks has become a main additional challenge. Prior work shows that polynomials can be used to fit Sigmoid function~\cite{aono2016scalable}, and high-degree polynomials can achieve very high accuracy~\cite{livni2014computational}. Hence, it is obvious that we can use the above method to implement the Sigmoid function, but high-degree polynomials will seriously reduce the efficiency, and using low-degree polynomials will lose a lot of accuracy.

\begin{figure}
  \begin{minipage}[t]{0.5\linewidth}
    \centering
    \includegraphics[width=0.8\linewidth]{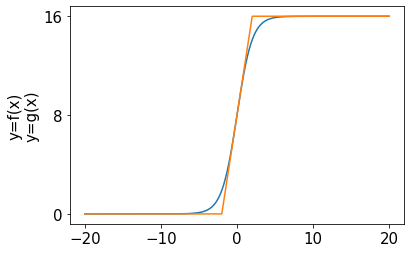}
    \caption{Our function $g(x)$}
    \label{ACF_fig}
  \end{minipage}%
  \begin{minipage}[t]{0.5\linewidth}
    \centering
    \includegraphics[width=1\linewidth]{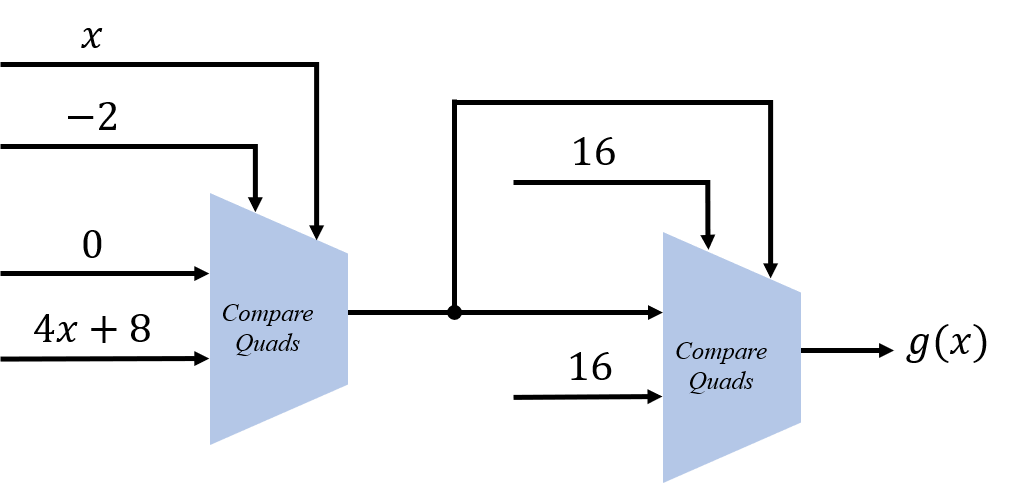}
    \caption{Constructure of our function}
    \label{COM_fig}
  \end{minipage}
\end{figure}

We borrow the idea in SecureML~\cite{mohassel2017secureml} which discusses that the piecewise function can also achieve high accuracy. Hence, we design a new MKTFHE-friendly activation function $g\left(x\right)$. In addition, in order to improve the accuracy, we fit the sigmoid function $f(x)$ in the form of the tangent at the origin, as shown in Fig.~\ref{ACF_fig}. Considering that MKTFHE only supports integers, we have zoomed the new activation function by 16 times. The function description is as follows:
$$
g(x) = \begin{cases}
	16,&x > 2 \\
	4x+8,&-2\leq x\leq 2\\
	0,&x < -2\\
\end{cases}
$$

Note that the comparison in our proposed activation function belongs to the second category of comparison, which is the comparison results are used for the next calculation instead of knowing the comparison results. Therefore, we can use two \emph{compare quads} in Subsection~\ref{com_qua} to implement the activation function following Fig.~\ref{COM_fig}. And details are shown in Algorithm~\ref{alg2}.



\begin{algorithm*}[t]
	\caption{New activation function}
	\label{alg2}
	\KwIn{MKTFHE parameter set $mkparams$, ciphertext $ct\left[l\right]=\mathsf{MKENC}\left(x,l\right)$ and the public keys of all participants $\{pk_i\}_k$}
	\KwOut{Ciphertext $c_{result}\left[l\right]=\mathsf{MKENC}\left(g\left(x\right),l\right)$}  
	\BlankLine
    Prepare the ciphertext $c\left(-2\right),c\left(16\right), c\left(0\right)$ and $c(4x+8)$\\
    Compute $\mathsf{Compare\_Quads}\left(ct\left[n\right],c\left(-2\right),c\left(0\right),c\left(4x+8\right)\right)$ to obtain the middle result $c_{mid}\left[l\right]$\\
    Compute $\mathsf{Compare\_Quads}\left(c_{mid}\left[l\right],c\left(16\right),c_{mid}\left[l\right],c\left(16\right)\right)$ to obtain the final result $c_{result}\left[l\right]$\\  
    Return $c_{result}\left[l\right]$
\end{algorithm*}

\subsection{Privacy-Preserving Machine Learning based on MKTFHE}
After we propose the new MKTFHE-friendly activation function, we will utilize it to compute the \emph{back propagation} in the logistic regression and neural networks. Considering the accuracy, we continue to use the Sigmoid function to calculate the partial derivative and maintain the structure of the iterative equation. And prior research also shows that if we change to compute the partial derivative of the linear activation function, the cross-entropy function is no longer convex, and the accuracy of training will incur more losses~\cite{mohassel2017secureml}.

In addition, since MKTFHE is for integers and Boolean, it is necessary to zoom the learning rate and other parameters into integers and modify the relevant calculation equation, so as to ensure that the model coefficients after training are also enlarged in proportion. 

We set the expansion factor $q$, use the integer learning rate $\alpha^\prime$ and the new iterative computation for logistic regression is below:
$$ h_\theta\left(x\right)=g(x) $$
$$\theta_j=\theta_jq-\alpha\prime\frac{1}{m}\sum_{i=1}^{m}{\left(h_\theta\left(x_i\right)-y_i\right)x_i^j}$$

There are $m$ neurons in the input layer, $n$ neurons in the hidden layer, and $p$ neurons in the output layer. Like the above logistic regression, we set the expansion factor $q$, use the integer learning rate $\alpha_1^\prime$, $\alpha_2^\prime$, $\beta_1^\prime$, $\beta_2^\prime$, and the rest definition is the same as Section~\ref{MLP}, the iterative computation for neural networks is below:
$$ o_i=g(\sum_{i=1}^{m} w_{ji} x_i^k q),{\hat{y}}_j^k=\sum_{i=1}^{n}{o_iv_{ji}q}$$

$$\Delta v_{ji}^{(n+1)}=\beta_2^\prime \Delta v_{ji}^{(n)}q-\alpha_1^\prime \Delta v_{ji}^{(n+1)}$$
$$\Delta w_{ij}^{\left(n+1\right)}=\beta_2^\prime\Delta w_{ij}^{\left(n\right)}+\left(q-\beta_2^\prime\right)o_i\left(q-o_i\right)x_j^k\sum_{j=1}^{p}{\left({\hat{y}}_j^k-y_j^kq\right)v_{ji}}$$
$$v_{ji}^{\left(n+1\right)}=v_{ji}^{\left(n\right)}q-\alpha_1^\prime\Delta v_{ji}^{\left(n+1\right)},{w_{ij}^{\left(n+1\right)}=w_{ij}^{\left(n\right)}q-\alpha_2^\prime\Delta w}_{ij}^{\left(n+1\right)}$$

\subsection{Our Framework}
After implementing the privacy-preserving logical regression and neural network training, we replace the original decryption with our proposed distributed decryption protocol and finally propose a distributed privacy-preserving machine learning framework based on MKTFHE, including four types of entities: participants, a cloud server, a CRS server, and a decryption party. The participants want to outsource computation and each of them holds their own part of data for model training which should not  be learned by a cloud server and other participants; The cloud servers are usually composed of one or more high-performance servers, which do not have their own data and only provide computing power. The CRS server is only responsible for generating the public parameters ( that is, common reference string) of the framework which can be included in the cloud server. The decryption party only joins in the distributed decryption and does not need too much computing power, which can be acted by a participant or a single server.

\begin{figure}[htbp]
	\centering
	\includegraphics[width=0.6\linewidth]{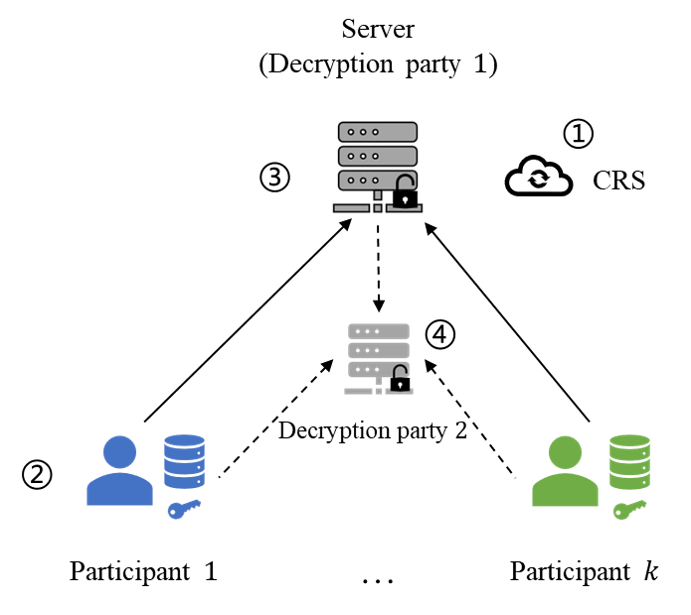}
	\caption{Framework of privacy-preserving machine learning}	
	\label{FLM}
\end{figure}

We take two participants as examples. The steps of the whole scheme are shown in Fig.~\ref{FLM}:
\begin{enumerate}[Step 1]
	\item Set up parameters: The CRS server call $\mathsf{MKTFHE.SETUP}(1^\lambda)$ to generate the $mkparams$ and communicates with the participants on the expansion factor $q$, then sends the set of parameters and factors $\{mkparams,q\}$ to each participant and the cloud server.
	\item Preprocess and encrypt data: Each participant uses the expansion factor $q$ to zoom or round the origin data and uses the $mkparams$ to call $\mathsf{MKTFHE.KEYGEN}\left(mkparams\right)$ to generate their own secret key and public key, then utilize their own secret key to encrypt the preprocessed data bit by bit by calling $\mathsf{MKTFHE.ENC}\left(m\right)$ and finally sends the public key and the ciphertext to the cloud server.
	\item Train ML models: The cloud server first expands the single key ciphertext from each participant to multi-key ciphertext, then uses the arithmetic operators like $\mathsf{MKADD}$, $\mathsf{MKSUB}$, $\mathsf{MKMUL}$, $\mathsf{MKDIV}$, and our newly designed \emph{homogenizer}, \emph{compare quads} to train the model. In the end, the cloud server sends the ciphertext model to the decryption party and all participants.
	\item Decrypt data: All the participants, the decryption party and the cloud server run the distributed decryption protocol together and the participants will get the plaintext in the end.
\end{enumerate}

\section{Implementation and Experiment}
\subsection{Implementation of Distributed Decryption Protocol}
\subsubsection{The Implementation and Experimental Environment}
Our code for the distributed decryption protocol is written in C + +, mainly using the arithmetic secret sharing of ABY~\cite{demmler2015aby} which is a very efficient two-party secure computing scheme. Considering that the final decryption only involves addition and subtraction without multiplication, this experiment neither needs to use oblivious transfer (OT) and other operations that require online interaction nor needs a semi-honest third party (STP) to assist in generating multiplication triples.

Noted that the ABY library only uses unsigned numbers and negative numbers cannot be directly computed. Therefore, we naturally regard the unsigned numbers as the number encoded by complement code and convert them into signed original code data after the final operation. In addition, since the finite field used in the MKTFHE scheme is a 32-bit torus, we straightly use the 32-bit arithmetic secret sharing in ABY library for computation.

The experiment of this subject runs on the Linux environment based on the following configuration.

\begin{enumerate}[(1)]
	\item Cloud server $S_0$ and $S_1$, is configured as Intel Xeon gold 5220 @2.2GHz processor, 256GB memory, and the operating system is Ubuntu 18.04 LTS;
	\item The client is Windows PC, Intel Core i7-8750H@2.20GHz processor, 16GB memory, and the operating system is Windows 10.
\end{enumerate}

In the experiment of LAN, we use two servers located in the same area. The network bandwidth is 512MB/s and the network delay is 0.35ms.

\subsubsection{Accuracy and Efficiency Analysis}

In this experiment, we compare with the original decryption scheme in MKTFHE and set the number of participants $k$ to 2, 4, and 8 respectively. We test in 10 groups for both decryption schemes and each group includes 1000 bits of ciphertext. Then we record the average time of decryption. Note that in the specific implementation, in the distributed decryption protocol, we use the SIMD technique in ABY library for parallel optimization to improve efficiency. The experimental results are shown below:

	

\begin{table*}[htb]
\caption{Implementation of distributed decryption} \label{tab1}
\begin{tabular*}{\hsize}{@{}@{\extracolsep{\fill}}cccc@{}}
                \toprule
			{Participants $k$} & {MKTFHE/s} & {Our protocol/s} & {Accuracy}\\
			\midrule
			2 & 0.024 & 0.261 & {100\%} \\
			4 & 0.050 & 0.268 & {100\%} \\
			8 & 0.112 & 0.263 & {100\%} \\
			\bottomrule

\end{tabular*}
    
\end{table*}

The result in Table \ref{tab1} shows that compared with the original MKTFHE, the efficiency of our scheme is relatively lower, but it’s still acceptable. We think the reason is mainly in the establishment of the secret sharing scheme and ciphertext transmission because the original scheme does not involve additional schemes and transmissions. By using the SIMD technique in ABY library, the decryption time basically remains unchanged with the increase of participants, while the decryption time of the original MKTFHE scheme increases linearly with the increase of participants. Therefore, in the case of multi-party participation, our scheme has more advantages in both security and efficiency.

\subsection{Implementation of Privacy Preserving Distributed Machine Learning}
\subsubsection{Data preprocessing}
Considering that MKTFHE only supports integers and Boolean, we need to preprocess the input data. We have two methods to preprocess, one is rounding and the other is zooming. The input data of logistic regression is in a rather large range while the input data of neural networks is relatively small, so we apply the rounding method on logistic regression and the zooming method on neural networks to keep the data precise. We store the zooming factor for the following computation to guarantee accuracy in an acceptable range.

\subsubsection{Implementation of Privacy-Preserving Logistic Regression}
The input data are generated by ourselves which are several sets of linear data with small random noise, and we mainly use 16-bit and 32-bit operators in this implementation.


We first use the 7-order Taylor polynomial (high enough order) formed Sigmoid function and our proposed activation function as the activation function in logistic regression to train the models with \emph{plaintext} of integer and float data. The result shows in Table~\ref{tab2} that in both integer and floating numbers, the accuracy of using a 7-order Taylor polynomial as an activation function is the highest, and using our proposed activation function can be close to that of a 7-order Taylor polynomial.

Then, we utilize the operators and other tools in MKTFHE to train the logistic regression models with the above different types of activation functions in \emph{ciphertext}. In addition to recording the accuracy and time in training in Table~\ref{tab3}, we also compare the computation time of different activation functions under MKTFHE in Table~\ref{tab4}. The result of the experiments shows that our scheme has no accuracy loss which means that the model trained in \emph{ciphertext} is the same as that in \emph{plaintext}, and the loss only occurs in the integer transfer stage. Using our proposed activation function can shorten the computing activation function time in \emph{ciphertext} by 10 times and significantly shorten the training time compared with the 7-order Taylor polynomial and the accuracy is close to it. Note that we also compare the 3-order Taylor polynomial with our proposed function in both \emph{plaintext} and \emph{ciphertext} and the result shows that comparing with the 3-order Taylor polynomial we can also shorten the computing activation time by 5 times with much better accuracy.


\begin{table*}[htb]
\caption{Logistic regression accuracy in plaintext} \label{tab2}
\begin{tabular*}{\hsize}{@{}@{\extracolsep{\fill}}cccc@{}}
            \toprule
			{Data type} & {7-order Taylor polynomial} & {3-order Taylor polynomial} & {Our function} \\
			\midrule
			Floating data & 98\% & 85\% & {95\%} \\
			Integer data & 95\% & 80\% & {92\%} \\
			\bottomrule    
\end{tabular*}
\end{table*}


\begin{table*}[htb]
\caption{Logistic regression accuracy in ciphertext} \label{tab3}
\begin{tabular*}{\hsize}{@{}@{\extracolsep{\fill}}ccc@{}}
            \toprule
			Activation function & Accuracy & Training time/iter/piece/s \\
			\midrule
			7-order Taylor polynomial & 95\% & 4049 \\
			3-order Taylor polynomial & 80\% & 2549 \\
            Our function & 92\% & 611 \\
			\bottomrule    
\end{tabular*}
\end{table*}


\begin{table*}[htb]
\caption{Computing activation function in ciphertext} \label{tab4}
\begin{tabular*}{\hsize}{@{}@{\extracolsep{\fill}}cccc@{}}
            \toprule
			& {7-order Taylor polynomial} & {3-order Taylor polynomial} & {Our function} \\
			\midrule
            Time/s & 1440 & 549 & 130 \\
			\bottomrule    
\end{tabular*}
\end{table*}

\subsubsection{Implementation of Privacy Preserving Neural Networks}
We use the Iris data set in sklearn~\cite{varoquaux2015scikit} as input data for neural networks, half of them for training and the rest for prediction. Like the above logistic regression, we also implement neural networks in both plaintext and ciphertext.

In \emph{plaintext}, we use the above different kinds of functions as activation functions to train the model with both integer data and floating data, and the same in \emph{ciphertext}. Note that we also compute every neuron in the same layer in parallel to optimize the code. The result of the experiments is shown in Table \ref{tab5}.

As the result shown in Table \ref{tab6}, using a 7-order Taylor polynomial as an activation function is more accurate and costs more time, but using our proposed activation function can greatly reduce the training time with close accuracy to it. Note that we also compare our function with a 3-order Taylor polynomial and the result shows that we shorten the training time with much better accuracy as well. 


\begin{table*}[htb]
\caption{Neural networks accuracy in plaintext} \label{tab5}
\begin{tabular*}{\hsize}{@{}@{\extracolsep{\fill}}cccc@{}}
            \toprule
			{Data type} & {7-order Taylor polynomial} & {3-order Taylor polynomial} & {Our function} \\
			\midrule
            Floating data & 96.23\% & 72.17\% & {95.46\%} \\
			Integer data & 94.67\% & 68.12\% & {94.15\%} \\
			\bottomrule    
\end{tabular*}
\end{table*}


\begin{table*}[htb]
\caption{Neural networks accuracy in ciphertext} \label{tab6}
\begin{tabular*}{\hsize}{@{}@{\extracolsep{\fill}}ccc@{}}
            \toprule
			Activation function & Accuracy & Training time/iter/piece/s \\
			\midrule
            7-order Taylor polynomial & 94.67\% & 7301 \\
 		3-order Taylor polynomial & 68.12\% & 6736 \\
 		Our function & 94.15\% & 4654 \\
			\bottomrule    
\end{tabular*}
\end{table*}

\subsubsection{Security analysis}
Our scheme is semantically secure under the (R)LWE assumption. We choose the same parameters in MKTFHE~\cite{MKTFHE, MKTFHE-op}. The achievement estimated security level of our scheme is 110-bit while the dimension of the TLWE problem is $k=1$.

\subsubsection{Distributed server cluster}
The training data, intermediate result e.g. descent gradient, and model parameters are all encrypted by single-key or multi-key and the evaluations among them are all fully homomorphic. So our scheme is "plug and play" for server clusters via most distributed machine learning strategies and typologies in~\cite{verbraeken2020survey} and does not leak any privacy in the distributed structure. 

\section{Conclusion and Discussion}
In this paper, we implement privacy-preserving logistic regression and neural networks with a distributed decryption protocol based on MKTFHE. Firstly, we introduce secret sharing to protect partial decryption and final decryption. Secondly, we design \emph{homogenizer} and \emph{compare quads} to implement our proposed MKTFHE-friendly activation function. Then, we utilize them to train privacy-preserving logistic regression and privacy-preserving neural networks. Finally, we formalize our distributed privacy-preserving machine learning framework. The experimental results show that the efficiency of our distributed decryption protocol is acceptable. Compared with using the Sigmoid function, the efficiency is greatly improved with our activation function and the accuracy is basically unchanged.

\section*{Acknowledgment}
This work is supported by National Natural Science Foundation of China (No. 62272131), Shenzhen Science and Technology Major Project (No. KJZD20230923114908017), 
and The Major Key Project of PCL (Grant No. PCL2023A05).

\bibliographystyle{plain}
\bibliography{easychair}

\begin{thebibliography}{10}

\bibitem{aono2016scalable}
Yoshinori Aono, Takuya Hayashi, Le~Trieu~Phong, and Lihua Wang.
\newblock Scalable and secure logistic regression via homomorphic encryption.
\newblock In {\em Proceedings of the Sixth ACM Conference on Data and Application Security and Privacy}, pages 142--144, 2016.

\bibitem{BoemerLCW19}
Fabian Boemer, Yixing Lao, Rosario Cammarota, and Casimir Wierzynski.
\newblock ngraph-he: a graph compiler for deep learning on homomorphically encrypted data.
\newblock In {\em Proceedings of the 16th {ACM} International Conference on Computing Frontiers, {CF} 2019}, pages 3--13, 2019.

\bibitem{BourseMMP18}
Florian Bourse, Michele Minelli, Matthias Minihold, and Pascal Paillier.
\newblock Fast homomorphic evaluation of deep discretized neural networks.
\newblock In {\em Advances in Cryptology - {CRYPTO} 2018}, volume 10993 of {\em Lecture Notes in Computer Science}, pages 483--512, 2018.

\bibitem{Bra16}
Zvika Brakerski and Renen Perlman.
\newblock Lattice-based fully dynamic multi-key {FHE} with short ciphertexts.
\newblock In {\em Advances in Cryptology - {CRYPTO} 2016}, volume 9814 of {\em Lecture Notes in Computer Science}, pages 190--213, 2016.

\bibitem{canetti2001universally}
Ran Canetti.
\newblock Universally composable security: A new paradigm for cryptographic protocols.
\newblock In {\em Proceedings 42nd IEEE Symposium on Foundations of Computer Science}, pages 136--145. IEEE, 2001.

\bibitem{MKTFHE}
Hao Chen, Ilaria Chillotti, and Yongsoo Song.
\newblock Multi-key homomorphic encryption from {TFHE}.
\newblock In {\em Advances in Cryptology - {ASIACRYPT} 2019 - 25th International Conference on the Theory and Application of Cryptology and Information Security}, volume 11922 of {\em Lecture Notes in Computer Science}, pages 446--472, 2019.

\bibitem{ChenGHHJLL18}
Hao Chen, Ran Gilad{-}Bachrach, Kyoohyung Han, Zhicong Huang, Amir Jalali, Kim Laine, and Kristin~E. Lauter.
\newblock Logistic regression over encrypted data from fully homomorphic encryption.
\newblock {\em {IACR} Cryptol. ePrint Arch.}, 2018:462, 2018.

\bibitem{CheonKKS18}
Jung~Hee Cheon, Duhyeong Kim, Yongdai Kim, and Yongsoo Song.
\newblock Ensemble method for privacy-preserving logistic regression based on homomorphic encryption.
\newblock {\em {IEEE} Access}, 6:46938--46948, 2018.

\bibitem{ChillottiGGI16}
Ilaria Chillotti, Nicolas Gama, Mariya Georgieva, and Malika Izabach{\`{e}}ne.
\newblock Faster fully homomorphic encryption: Bootstrapping in less than 0.1 seconds.
\newblock In {\em Advances in Cryptology - {ASIACRYPT} 2016}, volume 10031 of {\em Lecture Notes in Computer Science}, pages 3--33, 2016.

\bibitem{ChillottiGGI17}
Ilaria Chillotti, Nicolas Gama, Mariya Georgieva, and Malika Izabach{\`{e}}ne.
\newblock Faster packed homomorphic operations and efficient circuit bootstrapping for {TFHE}.
\newblock In {\em Advances in Cryptology - {ASIACRYPT} 2017}, volume 10624 of {\em Lecture Notes in Computer Science}, pages 377--408. Springer, 2017.

\bibitem{ChillottiGGI20}
Ilaria Chillotti, Nicolas Gama, Mariya Georgieva, and Malika Izabach{\`{e}}ne.
\newblock {TFHE:} fast fully homomorphic encryption over the torus.
\newblock {\em J. Cryptol.}, 33(1):34--91, 2020.

\bibitem{Clear15}
Michael Clear and Ciaran McGoldrick.
\newblock Multi-identity and multi-key leveled {FHE} from learning with errors.
\newblock In {\em Advances in Cryptology - {CRYPTO} 2015}, volume 9216 of {\em Lecture Notes in Computer Science}, pages 630--656, 2015.

\bibitem{CHET19}
Roshan Dathathri, Olli Saarikivi, Hao Chen, Kim Laine, Kristin~E. Lauter, Saeed Maleki, Madanlal Musuvathi, and Todd Mytkowicz.
\newblock {CHET:} an optimizing compiler for fully-homomorphic neural-network inferencing.
\newblock In {\em Proceedings of the 40th {ACM} {SIGPLAN} Conference on Programming Language Design and Implementation, {PLDI} 2019}, pages 142--156, 2019.

\bibitem{demmler2015aby}
Daniel Demmler, Thomas Schneider, and Michael Zohner.
\newblock Aby-a framework for efficient mixed-protocol secure two-party computation.
\newblock In {\em NDSS}, 2015.

\bibitem{CryptoNets16}
Ran Gilad{-}Bachrach, Nathan Dowlin, Kim Laine, Kristin~E. Lauter, Michael Naehrig, and John Wernsing.
\newblock Cryptonets: Applying neural networks to encrypted data with high throughput and accuracy.
\newblock In {\em Proceedings of the 33nd International Conference on Machine Learning, {ICML} 2016}, volume~48 of {\em {JMLR} Workshop and Conference Proceedings}, pages 201--210, 2016.

\bibitem{PMP4MLDS18}
Qi~Jia, Linke Guo, Zhanpeng Jin, and Yuguang Fang.
\newblock Preserving model privacy for machine learning in distributed systems.
\newblock {\em IEEE Transactions on Parallel and Distributed Systems}, 29(8):1808--1822, 2018.

\bibitem{MKTFHE-op}
Zoe~L Jiang, Jiajing Gu, Hongxiao Wang, Yulin Wu, Junbin Fang, Siu-Ming Yiu, Wenjian Luo, and Xuan Wang.
\newblock Privacy-preserving distributed machine learning made faster.
\newblock In {\em Proceedings of the 2023 Secure and Trustworthy Deep Learning Systems Workshop}, pages 1--14, 2023.

\bibitem{cryptoeprint:2018:1156}
Eunkyung Kim, Hyang-Sook Lee, and Jeongeun Park.
\newblock Towards round-optimal secure multiparty computations: Multikey fhe without a crs.
\newblock Cryptology ePrint Archive, Report 2018/1156, 2018.
\newblock \url{https://ia.cr/2018/1156}.

\bibitem{KimS0XJ18}
Miran Kim, Yongsoo Song, Shuang Wang, Yuhou Xia, and Xiaoqian Jiang.
\newblock Secure logistic regression based on homomorphic encryption.
\newblock {\em {IACR} Cryptol. ePrint Arch.}, 2018:74, 2018.

\bibitem{LeeP19}
Hyang{-}Sook Lee and Jeongeun Park.
\newblock On the security of multikey homomorphic encryption.
\newblock In {\em Cryptography and Coding - 17th {IMA} International Conference, {IMACC} 2019}, volume 11929 of {\em Lecture Notes in Computer Science}, pages 236--251, 2019.

\bibitem{livni2014computational}
Roi Livni, Shai Shalev-Shwartz, and Ohad Shamir.
\newblock On the computational efficiency of training neural networks.
\newblock {\em Advances in neural information processing systems}, 27, 2014.

\bibitem{Lopez12}
Adriana L{\'{o}}pez{-}Alt, Eran Tromer, and Vinod Vaikuntanathan.
\newblock On-the-fly multiparty computation on the cloud via multikey fully homomorphic encryption.
\newblock In {\em Proceedings of the 44th Symposium on Theory of Computing Conference, {STOC} 2012}, pages 1219--1234, 2012.

\bibitem{makri2017pics}
Eleftheria Makri, Dragos Rotaru, Nigel~P Smart, and Frederik Vercauteren.
\newblock Pics: Private image classification with svm.
\newblock {\em IACR Cryptol. ePrint Arch.}, 2017:1190, 2017.

\bibitem{mohassel2017secureml}
Payman Mohassel and Yupeng Zhang.
\newblock Secureml: A system for scalable privacy-preserving machine learning.
\newblock In {\em 2017 IEEE symposium on security and privacy (SP)}, pages 19--38. IEEE, 2017.

\bibitem{Mukherjee16}
Pratyay Mukherjee and Daniel Wichs.
\newblock Two round multiparty computation via multi-key {FHE}.
\newblock In {\em Advances in Cryptology - {EUROCRYPT} 2016}, volume 9666 of {\em Lecture Notes in Computer Science}, pages 735--763, 2016.

\bibitem{mukherjee2016two}
Pratyay Mukherjee and Daniel Wichs.
\newblock Two round multiparty computation via multi-key fhe.
\newblock In {\em Annual International Conference on the Theory and Applications of Cryptographic Techniques}, pages 735--763, 2016.

\bibitem{Peikert16}
Chris Peikert and Sina Shiehian.
\newblock Multi-key {FHE} from lwe, revisited.
\newblock In {\em Theory of Cryptography - 14th International Conference, {TCC} 2016-B}, volume 9986 of {\em Lecture Notes in Computer Science}, pages 217--238, 2016.

\bibitem{TianNYY21}
Yifan Tian, Laurent Njilla, Jiawei Yuan, and Shucheng Yu.
\newblock Low-latency privacy-preserving outsourcing of deep neural network inference.
\newblock {\em {IEEE} Internet Things J.}, 8(5):3300--3309, 2021.

\bibitem{varoquaux2015scikit}
Ga{\"e}l Varoquaux, Lars Buitinck, Gilles Louppe, Olivier Grisel, Fabian Pedregosa, and Andreas Mueller.
\newblock Scikit-learn: Machine learning without learning the machinery.
\newblock {\em GetMobile: Mobile Computing and Communications}, 19(1):29--33, 2015.

\bibitem{verbraeken2020survey}
Joost Verbraeken, Matthijs Wolting, Jonathan Katzy, Jeroen Kloppenburg, Tim Verbelen, and Jan~S Rellermeyer.
\newblock A survey on distributed machine learning.
\newblock {\em ACM Computing Surveys (CSUR)}, 53(2):1--33, 2020.

\end{thebibliography}

\appendix
\section{Attack on Existing Distributed Decryption Protocol}\label{Apx_1}
In this section, we propose a possible attack on~\cite{LeeP19} for the external passive adversary to obtain the final decryption result $m$. We show that the adversary only needs to collect the partial decryption broadcasted by each user to recover the message in the multi-key ciphertext.

As we mentioned above, the MKTLWE ciphertext is $({\bf a}_{1},\ldots, {\bf a}_{k}, b)$, which $b=\frac{1}{4}m-\sum^{k}_{i=1} \langle {\bf a}_{i}, {\bf s}_{i} \rangle + e$, $k$ is the number of total users, $m$ is one-bit message, and $e$ is an error of $b$.
In the partial decryption protocol of~\cite{LeeP19}, there are two kinds of partial decryption $p_{i, i}$ and $p_{i, j }$ both generated by user $u_i$:
\begin{itemize}
    \item $p_{i, i} = b + \langle {\bf a}_i, {\bf s}_i\rangle $ directly computed by its own secret key. Note that $p_{i, i}$ can also be viewed as a TLWE ciphertext containing the message $\mathsf{mess}$ encrypted by $j$-th user $u_j$ satisfying $p_{i, i} = \mathsf{mess} - \langle {\bf a}_j, {\bf s}_j \rangle + e$ and $\mathsf{mess} = \frac{1}{4}m - \sum_{t\ne i,j}^k \langle {\bf a}_t, {\bf s}_t \rangle$.
    \item $p_{i,j} = ({\bf a}_{i,j}, b_{i,j}) $ for each $j \in [k] \backslash \{i\}$ is the external product between the TLWE sample $({\bf a}_j, p_{i, i})$ and a trivial $\mathsf{TGSW}(1)$.
\end{itemize}
Then, user $u_i$ keeps the $p_{i, i}$ secret and broadcasts all the $p_{i,j} $ for each $j \in [k] \backslash \{i\}$. 

Note that the trivial $\mathsf{TGSW}(1) = {\bf Z}_t + {\bf H}$, each row of ${\bf Z}_t$ is $({\bf 0}, e)$ which is a trivial TLWE sample of zero with some noise. Thus, after the external product, ${\bf a}_{i,j} = {\bf a}_j$ and $b_{i,j}$ is still encrypted by user $u_j$ which still contains the message $\mathsf{mess}$. The only change is the noise $e$ becomes $\Tilde{e_j} = e + e_{add_j}$. So the $b_{i, j}$ satisfies the following equation:
$$b_{i,j}= \langle {\bf a}_j, {\bf s}_j \rangle + \mathsf{mess} + \Tilde{e_j} = \frac{1}{4}m - \sum_{t\ne i}^k \langle {\bf a}_t, {\bf s}_t \rangle + \Tilde{e_j}$$

So, the attack for the external adversary is that: after he collects the $p_{i,j}$ for some $j \ne i$ of all the users $u_i$ for $i\in [k]$, he can finish the final decryption to obtain $m$ on his own by computing $\sum_{i=1}^{k} p_{i, *} - (k-1) b$ and $*$ is denoted as some $j$ we don't care. For example, when $k=3$, one of the sets of $p_{i,j}$ supporting the attack is $p_{1,2}$, $p_{2,3}$, $p_{3,1}$.

{\bf Correctness of attack.} For $b = \frac{1}{4}m-\sum^{k}_{i=1} \langle {\bf a}_{i}, {\bf s}_{i} \rangle + e$, we have
\begin{align*}
    \sum_{i=1}^{k} p_{i, *}  &= k\cdot \frac{1}{4}m - k \cdot \sum_{t=1}^k \langle {\bf a}_t, {\bf s}_t \rangle + \sum_{i=1}^k \langle {\bf a}_i, {\bf s}_i \rangle + k\cdot \Tilde{e_*}\\
    &= k\cdot \frac{1}{4}m - (k-1) \cdot \sum_{t}^k \langle {\bf a}_t, {\bf s}_t \rangle + k\cdot \Tilde{e_*}
\end{align*}
$$\sum_{i=1}^{k} p_{i, *} - (k-1) b = \frac{1}{4}m + k \Tilde{e_*} - (k-1)e = \frac{1}{4}m + \Tilde{e}$$
If the magnitude of the error term $\Tilde{e}$ is less than $\frac{1}{8}$, the final decryption (attack) made by the adversary works correctly.

{\bf Error growth estimation.} Notice $e$ is the error in $b$ and $p_{i,i}$. After the external product, the noise becomes $\Tilde{e_*} = e + e_{add_*}$. We can estimate the magnitude of the growth $e_{add_*}$ from the external product noise propagation formula. So that we have 
$$\Tilde{e} = k \Tilde{e_*} - (k-1)e = k e + k e_{add_*} - (k-1)e = e + ke_{add_*}$$
$$\Vert \Tilde{e} \Vert_\infty \le \Vert {e} \Vert_\infty + k \max_*\{\Vert e_{add_*}\Vert_\infty\} = \Vert {e} \Vert_\infty + 2kl\beta  \max_*\{\Vert \mathsf{Err}({\bf A}_*) \Vert_\infty\} + 2\epsilon $$
To simplify, we omit some preliminaries only used in this section found in~\cite{LeeP19}, and some notions and symbols are the same as~\cite{LeeP19}. Therefore, the noise growth after the attack is quite small because $\mathsf{Err}({\bf A}_*)$ is the error of the  fresh TGSW ciphertext ${\bf A}_*$.

\end{document}